\documentclass[twoside,reqno]{bjp}
\usepackage{epsfig,cite}
\usepackage{graphics}
\usepackage{amssymb,amsmath}
\usepackage{times}
\begin{document}

\sloppy \raggedbottom

 \setcounter{page}{1}


\def\r{\mbox{\boldmath $r$}}
\def\R{\mbox{\boldmath $R$}}
\def\p{\mbox{\boldmath $p$}}
\def\q{\mbox{\boldmath $q$}}
\def\K{\mbox{\boldmath $K$}}
\def\diff{{\rm\,d}}                    
\newcommand{\inieq}{\begin{eqnarray}}            
\newcommand{\fineq}{\end{eqnarray}}            
\newcommand{\am}[1]{\textbf{#1 (*AM*)}}
\newcommand{\eep} { $(e,e^{\,\prime}p)$ } 
\def\ee{\mbox{$\left(e,e^{\prime}\right)$\ }}
\def\ie{i.e.}
\def\dblone{\hbox{$1\hskip -1.2pt\vrule depth 0pt height 1.6ex width 0.7pt
     \vrule depth 0pt height 0.3pt width 0.12em$}}
     \def\chiral4lo{$\mathrm{N}^4\mathrm{LO}$}


\title{Phenomenological and Theoretical Optical Potentials\thanks{Contribution presented at the Workshop ``Advanced Aspects of Nuclear Structure and Reactions at Different Energy Scales", 25-28 April 2017, Arbanasi, Bulgaria.}}

\runningheads{Phenomenological and Theoretical Optical Potentials}{C.~Giusti}

\begin{start}
\author{C.~Giusti}{}

\address{Dipartimento di Fisica Nucleare e Teorica, 
Universit\`a degli Studi Pavia, Pavia, and INFN, Sezione di Pavia,  Via Bassi 6, I-27100 Pavia, Italy}{}




\begin{Abstract}
The optical potential (OP) is a powerful instrument for calculations 
on a wide variety of nuclear reactions, in particular, for quasi-elastic lepton-nucleus scattering. Phenomenological OPs are successful in the description of data but may produce uncertainties 
in the interpretation of the results. Two recent theoretical OPs are presented: a global relativistic folding OP, that has been employed in relativistic models for quasi-elastic lepton-nucleus scattering, and a nonrelativistic OP derived from nucleon-nucleon chiral potentials at fourth order (N$^4$LO), that has been applied to elastic proton-nucleus scattering.
\end{Abstract}

\end{start}

\section[]{Introduction}

The Optical Potential (OP) describes the nucleon-nucleus ($NA$) interaction in elastic scattering \cite{Hodg}. Its use can be extended to inelastic scattering and to calculate the cross section of a wide variety of nuclear reactions. In models for Quasi-Elastic (QE) electron and neutrino-nucleus scattering the OP describes the Final-State Interaction (FSI) between the outgoing nucleon (that is emitted in QE scattering) and the residual nucleus~\cite{book,eenr,eesym,ee,eep,cc,nc,eea,confee,eepex,ex,confcc,compmini,prd,compnc}.
The OP can be obtained phenomenologically, by assuming a form of the potential and a dependence by a number of adjustable parameters for the real and imaginary parts that characterize the shape of the nuclear density distribution and that vary with the nucleon energy and the nucleus mass number. These parameters are adjusted to optimize the fit to elastic proton-nucleus ($pA$) scattering data. Alternatively and more fundamentally, the OP can be obtained microscopically. The calculation requires, in principle, the solution of the full many-body nuclear problem, which is beyond present capabilities. In practice, some approximations must necessarily be adopted to make calculations feasible. In general we do not expect that microscopic OPs are able to describe available experimental data better than phenomenological OPs, but they should have more theoretical content and a greater predictive power when applied to situations where data are not yet available.

The use of phenomenological OPs in calculations of QE lepton-nucleus scattering makes the results generally successful in comparison with data. The availability of different phenomenological OPs, however,  may produce theoretical uncertainties on the numerical predictions of the models and ambiguities in the interpretaion of the results. Thus the need arises to obtain more theoretical OPs. 

A new OP has been built within the relativistic impulse approximation from a folding approach \cite{GRFOP} and it has been employed in calculations of QE lepton-nucleus scattering. 
A new microscopic OP has been obtained for elastic $NA$ scattering from Nucleon-Nucleon ($NN$) chiral potentials \cite{ChOP}.

In the following the role of the OP in models for QE lepton-nucleus scattering is discussed. Then, recently derived theoretical OPs are presented.   

\section{Optical Potential in Quasi-Elastic Lepton-Nucleus Scattering}

In the QE kinematic region the nuclear response to an electroweak probe is
dominated by the process of one-nucleon knockout, where the probe interacts with only one nucleon of the target which is then emitted by a direct one-step knockout mechanism. In electron scattering experiments
the emitted nucleon can be detected in coincidence with the scattered electron and the residual nucleus is left in a bound or continuum state.
The exclusive \eep knockout reaction for transitions to discrete bound eigenstates of the residual nucleus has been widely investigated \cite{book}.
If only the scattered electron is detected, the final nuclear state is not determined and the measured cross section includes all the available final states. This is the inclusive \ee scattering.

Lepton-nucleus scattering is usually described in the 
one-boson exchange approximation, where the cross section is obtained from 
the contraction between the lepton tensor, which 
essentially depends only on the lepton 
kinematics,  and the hadron tensor, whose components 
are given by products of the matrix elements of the nuclear current between 
the initial and final nuclear states. 

In the QE region, electron scattering can be described in the impulse approximation (IA). The IA assumes: for the exclusive scattering that the interaction occurs through a one-body current only with the 
quasi-free ejectile nucleon; for the inclusive scattering that the cross section is given by the incoherent sum, over all the target nucleons, of integrated one-nucleon knockout processes. Then, we must describe the FSI between the ejectile and the residual nucleus.

In the exclusive \eep reaction FSI is usually described in the distorted-wave 
IA (DWIA) by a complex OP, where the imaginary part gives an absorption that reduces the calculated cross section. This reduction is essential to reproduce \eep data. Models based on a nonrelativistic
DWIA or a relativistic RDWIA are indeed able to give an excellent description 
of \eep data~\cite{book,eep}. 

In the inclusive scattering a model based on the DWIA, where the 
cross section is given by the sum of all integrated one-nucleon knockout 
processes and FSI are described by an OP with an 
imaginary absorptive part, is conceptually wrong. The OP describes elastic
$NA$ scattering and its imaginary part accounts for the fact that, if other 
channels are open besides the elastic one, part of the incident flux is lost in 
the elastically scattered beam and goes to the inelastic channels which are 
open. In the exclusive reaction only one channel is considered and it 
is correct to account for the flux lost in the selected channel.  In the 
inclusive scattering all the final-state channels are included, the flux lost 
in a channel must be recovered in the other channels, and in the sum over all 
the channels the flux can be redistributed but must be conserved. In every channel flux is lost toward other channels and flux is gained due to the flux lost in the other channels toward that channel. In the DWIA flux is lost in every channel. 

For the inclusive scattering a different but consistent model was developed to describe FSI: the nonrelativistic or relativistic Green's Function (GF or RGF) model \cite{eenr,eesym,ee}. The model is still based on the IA: the probe interacts through a one-body current with an ejectile nucleon, a sum is performed over all the nucleons of the target and FSIs are described by the same complex and energy dependent OP as in the exclusive scattering. The formalism, however, translates the flux lost to inelastic channels, represented by the  imaginary part of the OP, into the strength observed in the inclusive reaction. 
In the model the components of the hadron tensor are written in terms of the single-particle
(s.p.) optical model Green's function. The explicit calculation of the s.p. 
Green's function can be avoided exploiting its spectral 
representation, which is based on a biorthogonal expansion in terms of the 
eigenfunctions of the non-Hermitian OP and of its Hermitian conjugate.  
The s.p. expression of the hadron-tensor 
components is then obtained in a form  which contains matrix elements of the 
same type as the DWIA ones of the exclusive \eep process, but these matrix 
elements now involve eigenfunctions of the
OP and of its Hermitian conjugate, where the opposite sign
of the imaginary part gives in one case an absorption and
in the other case a gain of strength. Therefore, in the model 
the imaginary part redistributes the flux lost in every channel in the
other (inelastic) channels and in the sum over all the channels the total
flux is conserved. With the use of a complex OP the model can recover 
contributions of inelastic channels that are not included in usual models based on the RIA: all the available final-state channels are included, not only direct one-nucleon emission processes.
The energy dependence of the OP reflects the different contribution of the different inelastic channels that are open at different energies and makes the results very sensitive to the kinematic conditions of the calculation.

The model has been applied to the inclusive QE \ee reaction \cite{eenr,eesym,ee,confee}, the RGF has been extended to Charge-Current QE (CCQE)\cite{cc,confcc,compmini} and Neutral Current Elastic (NCE) \cite{nc,prd,compnc} (anti)neutrino-nucleus scattering. 
Calculations are usually performed adopting phenomenological OPs, with parameters adjusted to optimize the fit to elastic $pA$  scattering data.

The RGF results give a good description of the experimental \ee cross sections, in particular, in kinematic situations where the longitudinal response is 
dominant~\cite{ee,confee,ex} and are able to describe CCQE and NCE MiniBooNE data and CCQE Miner$\nu$a data~\cite{prd,compmini}. In comparison with the MiniBooNE data, the RGF results are usually larger than the 
results of other models based on the IA, which, in general, underpredict data. The enhancement can be ascribed to the effect of the inelastic channels, which are recovered by the imaginary part of the OP and that are not included in other models based on the IA. 

The model is therefore generally successful, but there are some caveats due to the use of phenomenological OPs. The imaginary part of the OP recovers and includes contributions beyond direct one-nucleon knockout, such as, for instance, rescattering of the outgoing nucleon and some multi-nucleon and non-nucleonic processes in the final state. A phenomenological OP, however, does not allow us to disentangle and evaluate the role of a specific inleastic contribution. 
Phenomenological OPs are obtained through a fit to elastic $pA$ scattering data. These data, however, do not completely constrain the shape  and the size of the OP. Different phenomenological OPs are available, they are able to give an equivalently good description of elastic $pA$ data, but they are different, in particular, their imaginary parts are different and can give different inelastic contributions and therefore different results. In many cases the differences are small or even negligible, but there are also situations where the differences are large and produce theoretical uncertainties on the predictions of the models.

\section{Global Relativistic Folding Optical Potential}

To reduce the uncertainties produced by phenomenological OPs and to ascertain to what extent  the RGF predictions can be relied upon,  the need arises to build microscopic relativistic OPs (ROPs). A new ROP has been built for $^{12}$C, a nucleus that is often used in neutrino-scattering experiments.  The new ROP is global, just like the phenomenological ROPs used in previous RGF calculations, \ie,   spanning a large range of kinetic energies of the nucleon, and it has been built from a  folding approach \cite{GRFOP}. The shape of the potential is constrained by the assumed shape of the nuclear density and the strength of the different contributions is essentially dictated by their respective contents in the effective parametrization of the $NN$ scattering amplitudes. Indeed, within the RIA, one can build OPs to study
$NA$ reactions which provide excellent quantitative descriptions
of elastic proton-scattering observables from various
spin-saturated spherical nuclei~\cite{HLF,HLFa}. Two basic ingredients underly
the realization of these folding potentials: a suitable analytical
representation of the $NN$-interaction and an appropriate relativistic model of
nuclear densities. The GRFOP has been
generated by folding the Horowitz-Love-Franey (HLF)  $t$-matrix with the
relevant relativistic  mean-field Lorentz densities via the so-called $t\rho$-approximation. 

In comparison with the phenomenological OPs the new GRFOP: 1) is derived from all available experimental data of elastic
proton scattering on $^{12}$C we are aware of; 2) stems from a folding approach, with neutron density fitted to data and proton
density taken from electron-scattering experiments;
3) the same nuclear densities are used at all the energies in the range between  $20$ and  $1040$ MeV;
4) the imaginary term is built from the effective $NN$ interaction.

The GRFOP reproduces the energy dependence of the 
experimental cross sections for the elastic proton scattering 
on $^{12}$C in the energy range  between 20 and 1040 MeV \cite{GRFOP}.  The agreement with  the experimental analyzing power is comparable \cite{GRFOP} to the one obtained with the phenomenological EDAI and EDAD1 potentials \cite{chc}, which have been widely used in RDWIA and RGF calculations. %
 

\begin{figure}[tb]
\epsfig{file=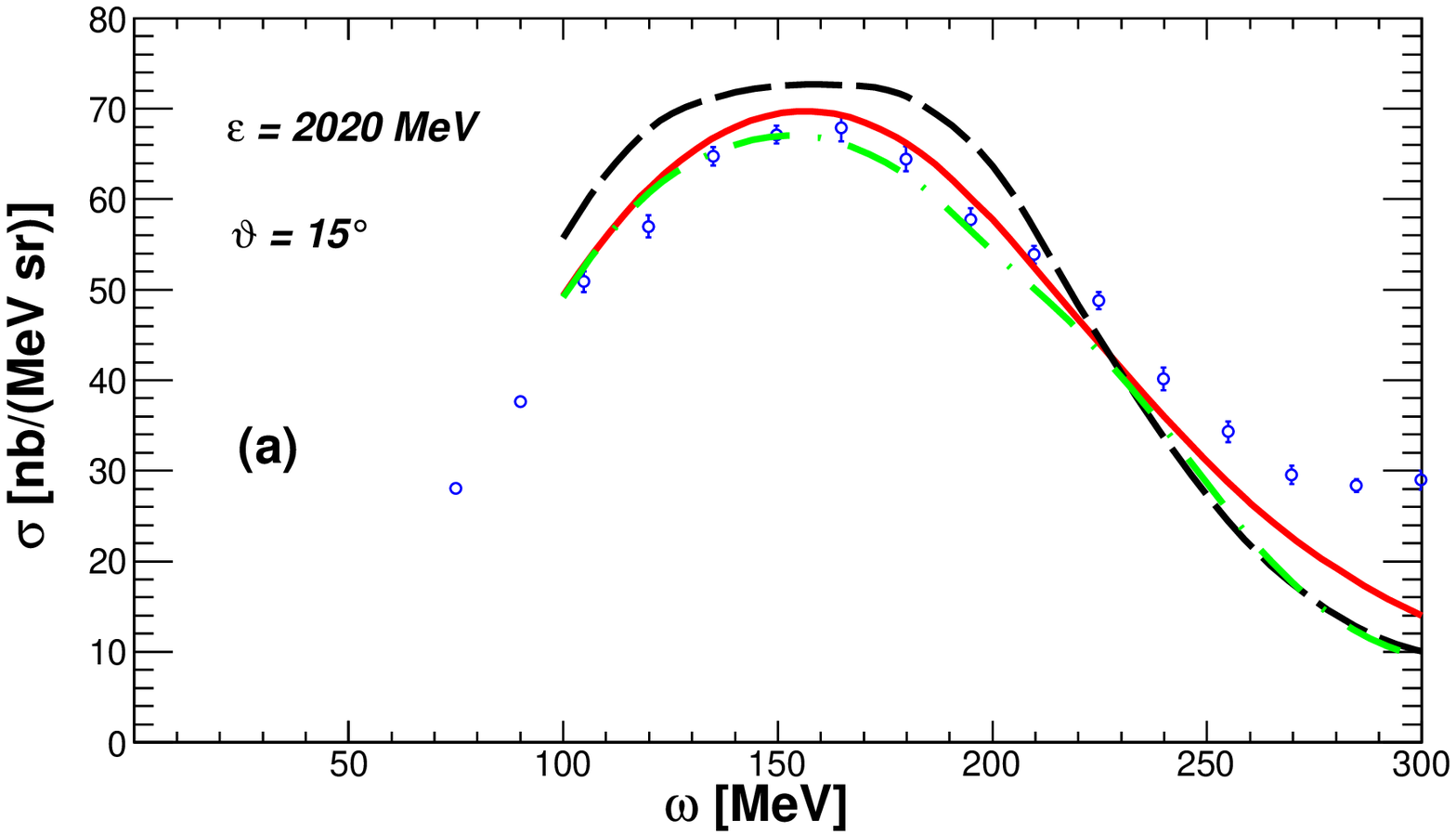,width=65mm}
\epsfig{file=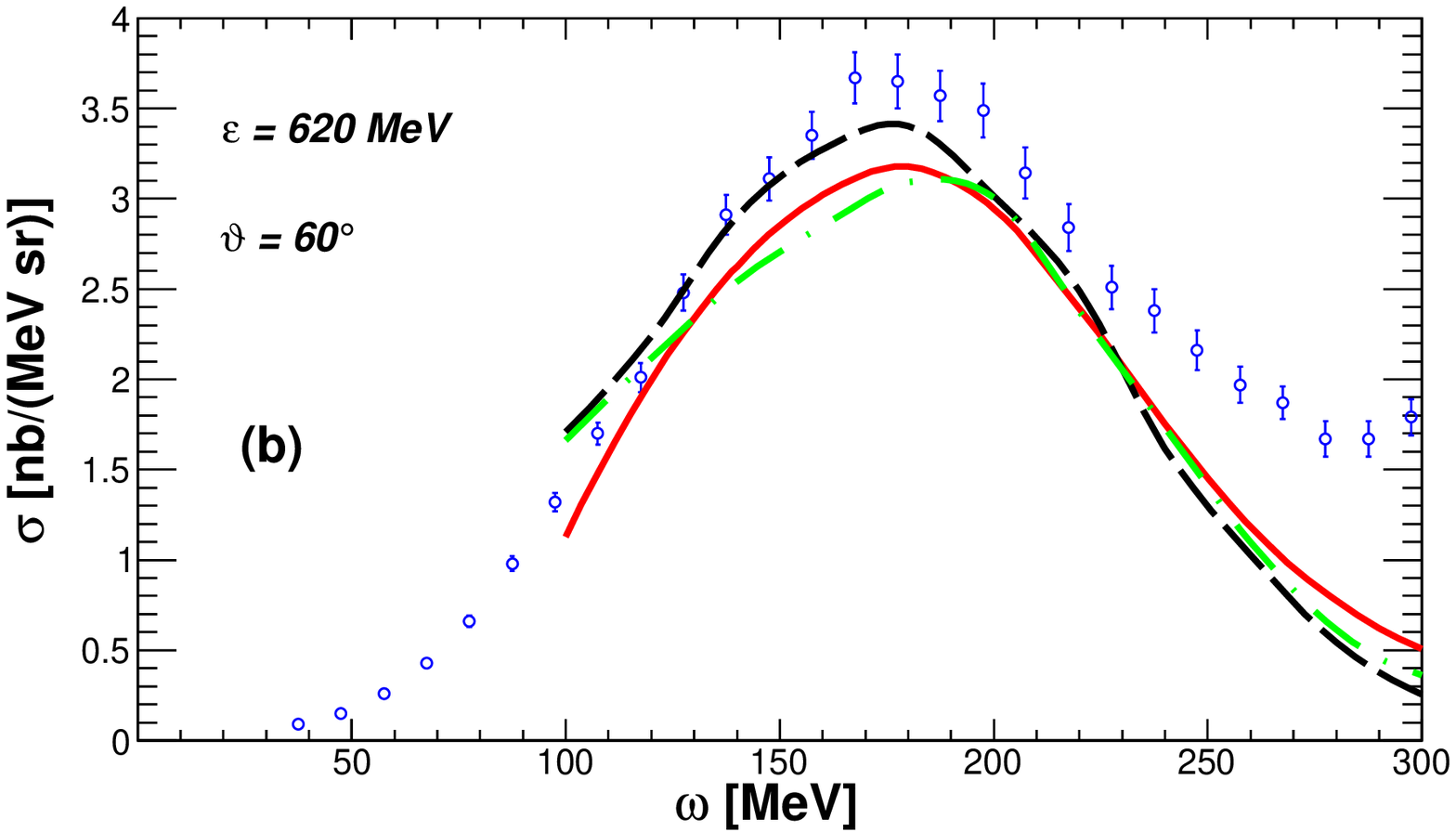,width=65mm} 
\caption{Differential cross section of the $^{12}$C$(e,e^{\prime})$ reaction for different beam energies and electron scattering angles.  Line convention: RGF-GRFOP (green), RGF-EDAI (black), RGF-EDAD1 (red). Experimental data from  \cite{ee1,ee2}.
\label{eeinc}}
\end{figure}
\begin{figure}[tb]
\epsfig{file=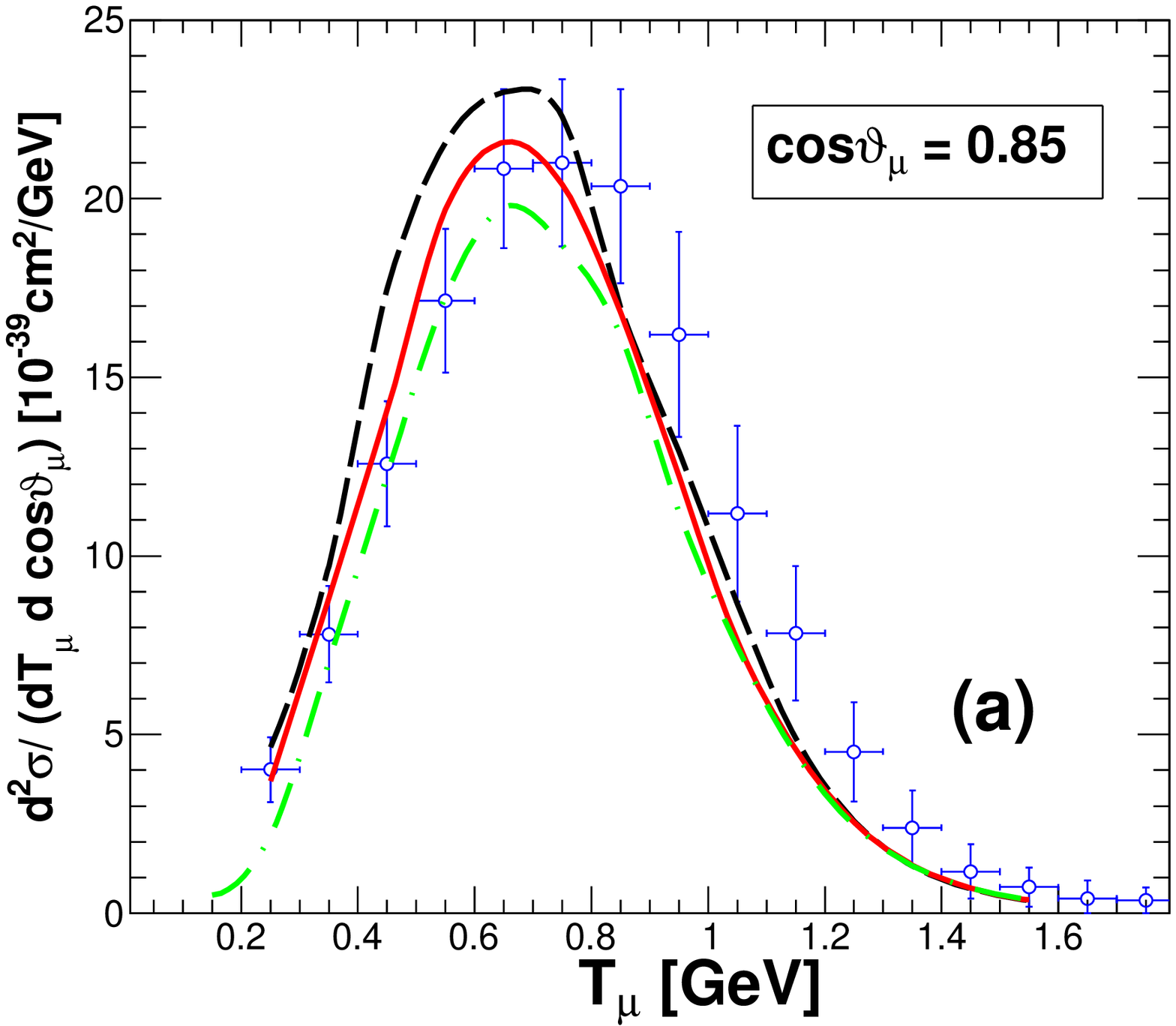,width=60mm} 
\epsfig{file=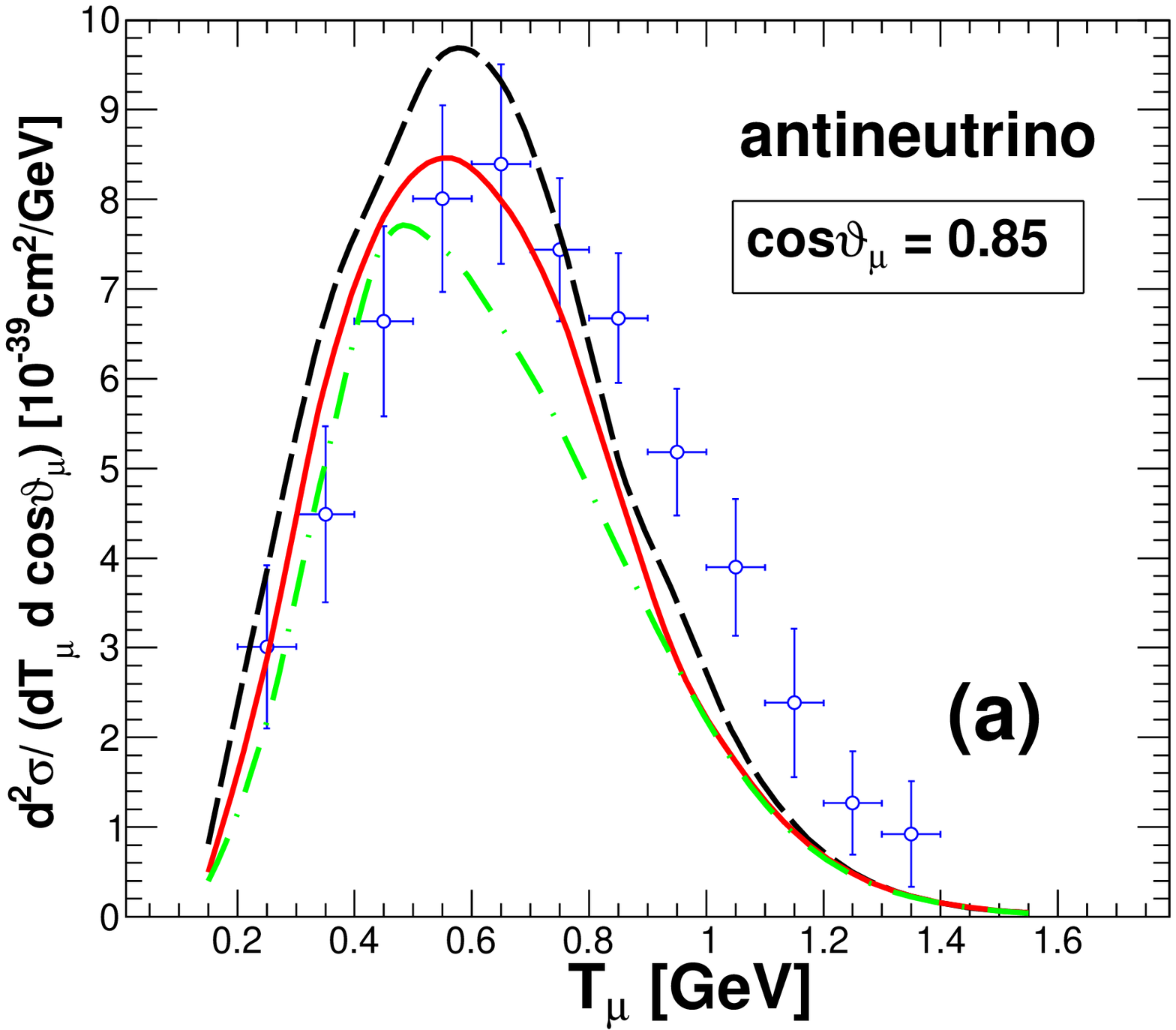,width=60mm}\\   
\epsfig{file=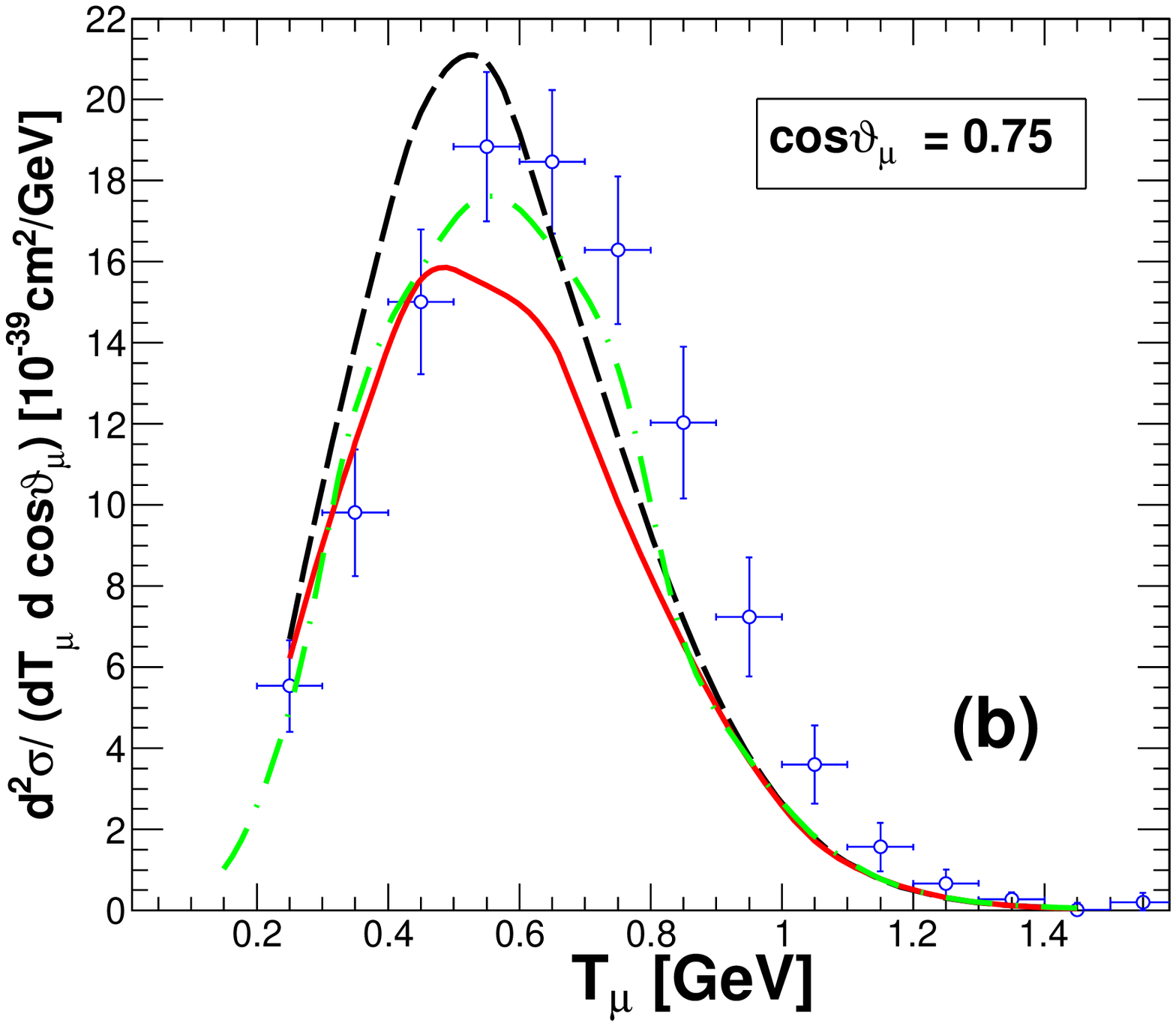,width=60mm} 
\epsfig{file=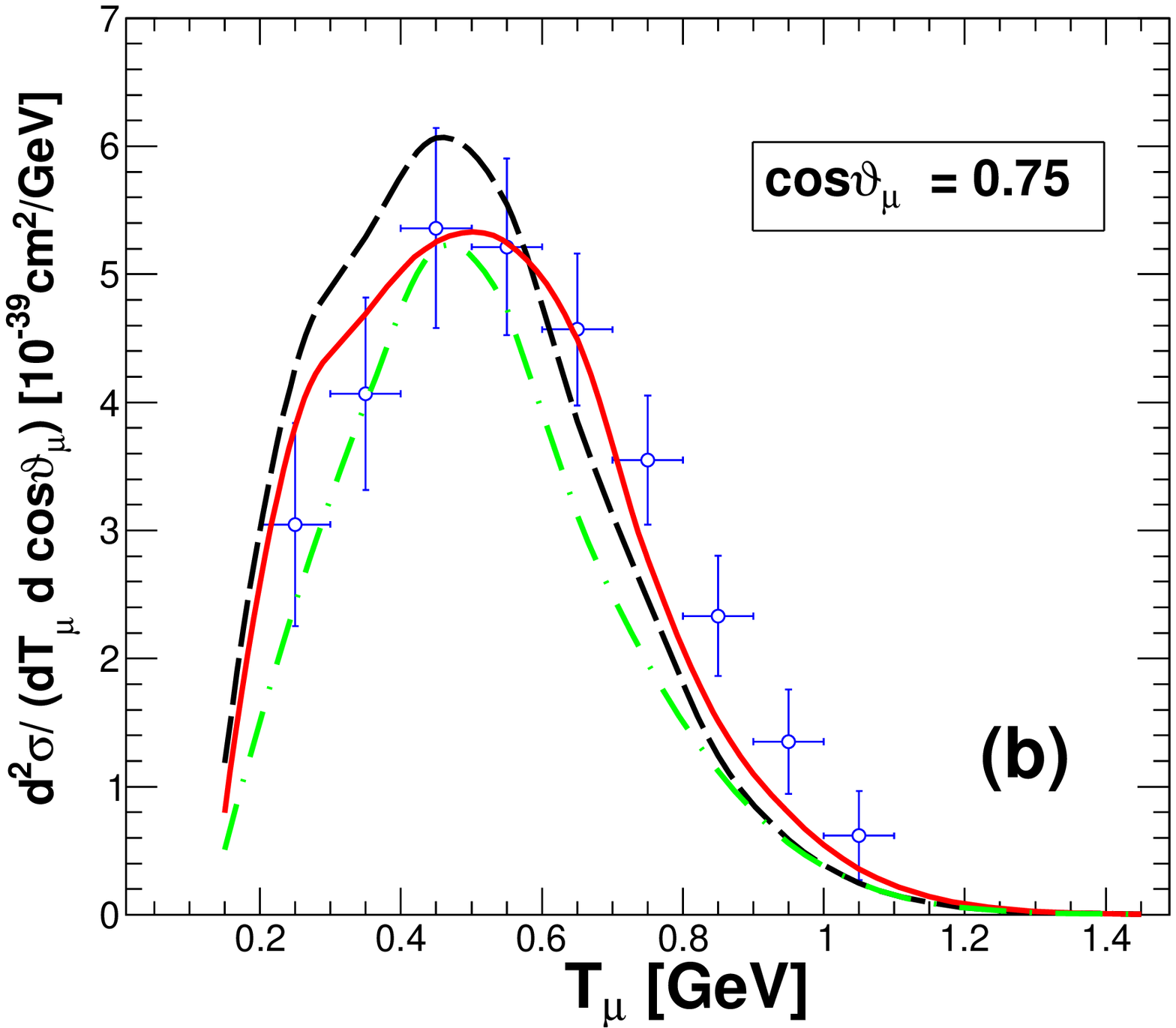,width=60mm}\\  
\epsfig{file=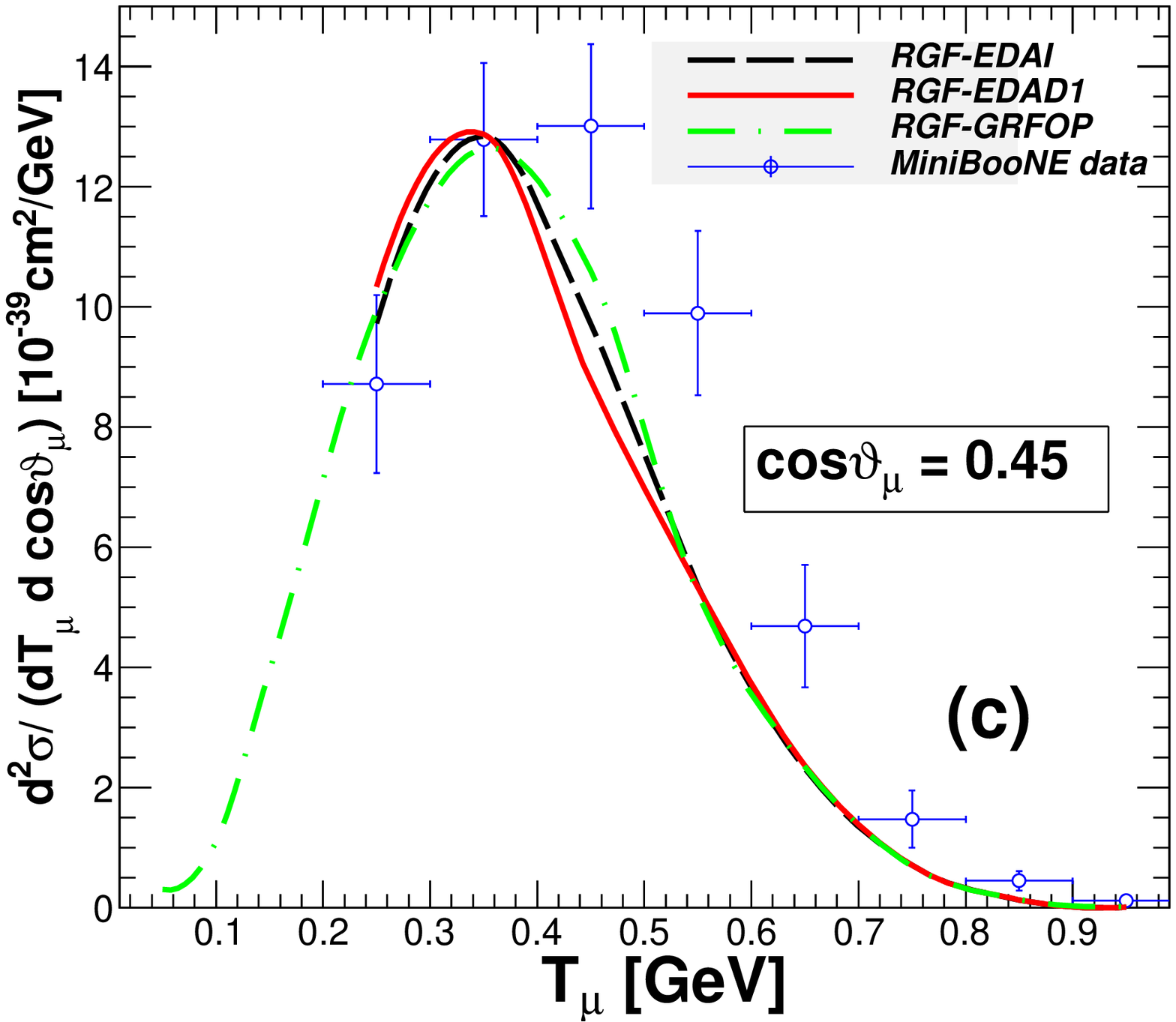,width=60mm} 
\epsfig{file=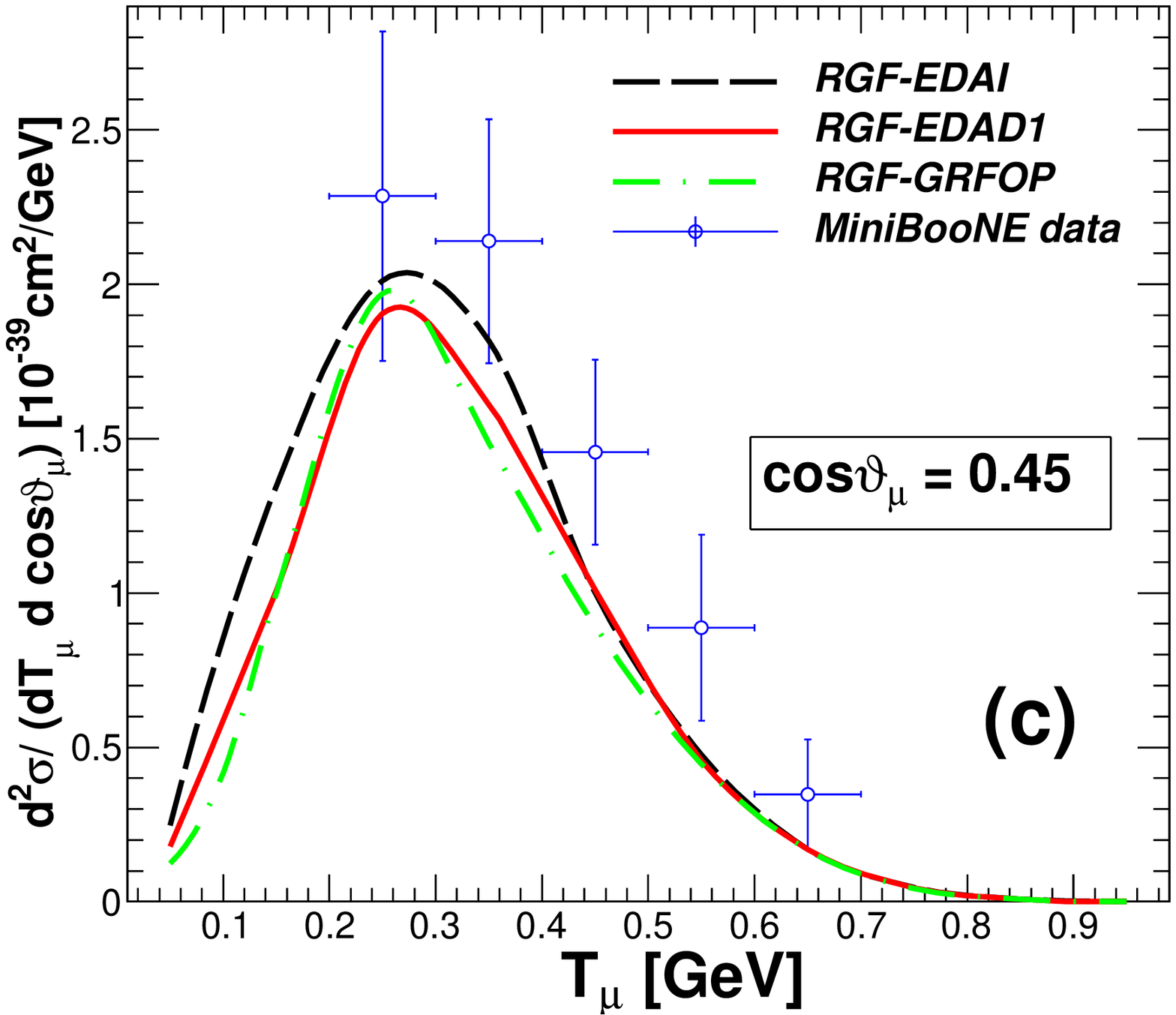,width=60mm} 
\caption{Flux-averaged double-differential cross section per target nucleon for the CCQE $^{12}$C$(\nu_{\mu} , \mu^-)$ (left panels) and 
$^{12}$C$(\bar\nu_{\mu} , \mu^-)$ (right panels) reactions as a function of the muon kinetic energy $T_{\mu}$ for three bins of the muon scattering angle $\cos\vartheta_\mu$. Line convention as in Figure~1. Experimental data from MiniBooNE \cite{miniboone, minibooneanti}.\label{ddneu}}
\end{figure}
The GRFOP has been tested within the RGF for QE electron scattering. The results are in generally good agreement with the experimental \ee cross sections  and close to the results obtained with EDAI and EDAD1 \cite{GRFOP}. 
An example is shown in Figure~1, where the results of the three ROPs are compared with the experimental $^{12}$C$(e,e^{\prime})$ differential cross sections for two different kinematics.  The agreement with data is generally satisfactory. The differences among the three RGF results are qualitatively similar in the two kinematics, where the momentum transfer in the QE peak region is approximately the same, \emph{i.e.} $q\approx 0.55$ GeV$/c$: the RGF-EDAI cross section is larger than the RGF-EDAD1 and  RGF-GRFOP ones. 
The experimental cross section in panel (a) is well described  in the peak region by RGF-EDAD1 and RGF-GRFOP and slightly overpredicted by RGF-EDAI; in panel (b) it is slightly underpredicted by all the calculations. 

The comparison of the RGF results with CCQE MiniBooNE data is presented in Figure~2, where the flux-averaged double-differential cross sections per target nucleon for $\nu$ and $\bar\nu$ scattering are plotted as functions of the muon kinetic energy $T_{\mu}$ for three bins of the muon scattering 
angle $\vartheta_{\mu}$.
A good agreement with the shape of the experimental cross sections is generally obtained with all the three ROPs.
The RGF-GRFOP results lie, in general, between the RGF-EDAI and RGF-EDA1 ones
and are in comparable or even better agreement with the
data. 

The GRFOP reduces the uncertainties in the 
predictions of the RGF model and confirms our 
previous findings in comparison with the data.
The RIA can provide successful ROPs with similar fits to elastic $NA$ scattering data and that can be considered as a useful alternative to phenomenological OPs.  

\section{Theoretical Optical Potential Derived from $NN$ Chiral Potentials}

\begin{figure}[t]
\centering{\epsfig{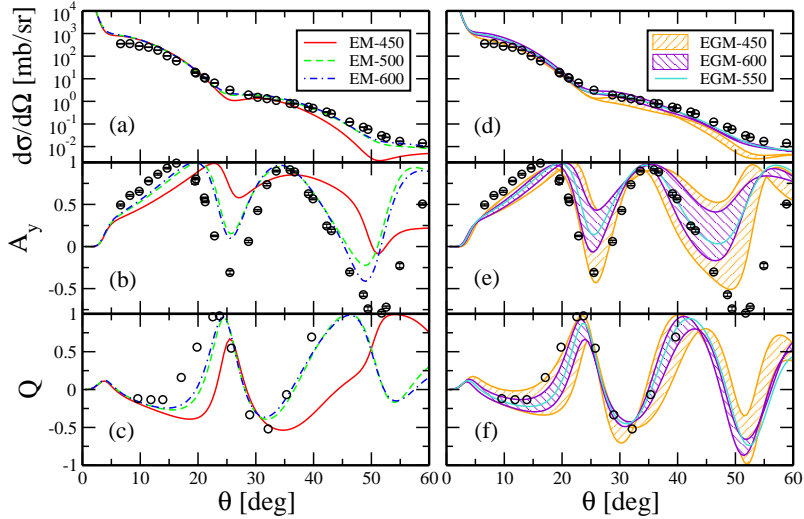}}
\caption{Scattering observables as a function of the center-of-mass scattering angle $\theta$  for elastic proton scattering on $^{16}$O computed at $200$ MeV (laboratory energy) with EM potentials \cite{macn3lo} (left panel) and EGM potentials \cite{epen3lo} (right panel). 
All potentials are denoted by the value of the LS cutoff. Bands in the left panel are produced by changing the $\tilde{\Lambda}$ cutoff.
Data from \cite{kelly,exfor}
\label{f1-ani}}
\end{figure}
Recently, a new microscopic optical potential for elastic $pA$ scattering has 
been obtained \cite{ChOP} employing two-body chiral potentials. The theoretical framework basically follows the approach of \cite{KMT} where the Watson 
multiple-scattering theory was developed expressing the $NA$ OP by a series expansion in terms of the free $NN$ scattering amplitudes.

Elastic $pA$ scattering can be formulated in the 
momentum space by the full Lippmann-Schwinger (LS) equation \cite{Hodg}
\begin{equation}\label{generalscatteq}
T = V \left( 1 + G_0 (E) T \right)\, , 
\end{equation}
where $V$ is the external interaction which, if we assume only two-body forces, is given by the sum over all the target nucleons  of two-body potentials describing the interaction of each target nucleon with the incident proton and
$G_0 (E)$ is the free Green's function for the $(A+1)$-nucleon system.

As a standard procedure, Eq.~(\ref{generalscatteq})
is separated into a set of two coupled integral equations: the first one for the so-called $T$ matrix
\begin{equation}\label{firsttamp}
T = U \left(1+ G_0 (E) P T\right) \, 
\end{equation}
and the second one for the optical potential $U$
\begin{equation}\label{optpoteq}
U = V \left(1+ G_0 (E) Q U \right)\, .
\end{equation}
The operator $P$ projects onto the elastic channel and $Q$ is defined  by the completeness relation $P + Q = \dblone$.

A consistent framework to compute $U$ and $T$ is provided by the spectator expansion of the nonrelativistic multiple-scattering theory, retaining only the first-order term, corresponding to the single-scattering approximation, where only one target-nucleon interacts with the projectile. In addition, we adopt the impulse approximation, where nuclear binding on the interacting target nucleon is neglected.
After some lenghty manipulations \cite{ChOP}, the OP is obtained in the so-called optimum factorization approximation as the product of the free
$NN$ $t$ matrix and the nuclear matter densities
\begin{equation}
\label{optimumfact}
U ({\q},{\K};\omega) = \frac{A-1}{A} \, \eta ({\q},{\K}) 
 \sum_{N = n,p} t_{pN} \left({\q}, {\K}, \omega \right) \, \rho_N (q) \, ,
\end{equation}
where ${\q}$ and ${\K}$ are the momentum transfer and the total 
momentum, respectively, in the $NA$ reference frame, 
$t_{pN}$ represents the proton-proton ({\it pp}) and proton-neutron ({\it pn}) 
$t$ matrix, $\rho_N$ the neutron and proton profile density,
and $\eta ({\q},{\K})$ is the M\o ller factor, that imposes the 
Lorentz invariance of the flux when we pass from the NA to the NN
frame in which the $t$ matrices are evaluated.  
Through the dependence of $\eta$ and $t_{pN}$ upon ${\K}$, the optimally factorized OP given in Eq.~(\ref{optimumfact}) exhibits nonlocality and off-shell effects. 
The energy $\omega$ 
is fixed at one half of the kinetic
energy of the projectile in the laboratory system.

The optimally factorized OP is then written 
exploiting its spin-dependent component \cite{ChOP} and then expanded 
on its partial-wave components. Once the $LJ$ components of the elastic transition operator are determined,
the calculation of the scattering observables (the unpolarized differential cross section ${\rm d}\sigma/{\rm d}\Omega$, the analyzing power $A_y$, and the spin rotation $Q$) is straightforward.

The calculation of the OP requires two basic ingredients: the $NN$ potential and the nuclear 
densities. The latter quantities are computed within the relativistic mean field  description \cite{Nik1} of spherical nuclei using a density-dependent meson-exchange model, where the couplings between mesonic and baryonic fields are assumed as function of the density itself \cite{Nik2}.
For the $NN$ potential OP models have always employed ``realistic" $NN$ potentials, which are able to reproduce the experimental $NN$ phase shifts with a $\chi^2$/datum $\simeq 1$. In \cite{ChOP} we have adopted 
chiral potentials, with the purpose to study the domain of applicability of microscopic two-body 
chiral potentials in the construction of an OP. Two different versions of chiral potentials at fourth order N$^3$LO in the chiral expansion parameter have been used, developed by Entem and Machleidt (EM) \cite{macn3lo} and Epelbaum, Gl\"ockle, and Mei\ss ner (EGM) \cite{epen3lo}. 
Both versions employed a regulator function $f_\Lambda$ (with three choices of the cutoff: $\Lambda=$ 450, 600, and 500 (EM) or 550 (EGM) MeV) to regulate the high-momentum components in the LS equation, but they approached differently the treatment of the short-range part of the two-Pion Exchange (2PE) contribution, that has unphysically strong attraction.
EM treated divergent terms in the 2PE contributions with dimensional regularization (DR), while EGM used a spectral function regularization (SFR), which introduces an additional cutoff $\tilde{\Lambda}$ in the evaluation of 
the potential and, as a consequence,  also into the perturbative resummation.

The results produced by the two different versions of the chiral potential for different LS cutoffs have been compared for the $NN$ scattering amplitudes and for the observables of elastic proton scattering on $^{16}$O \cite{ChOP}. 
The agreement with data gets worse increasing the energy.
At 100 MeV all the $NN$ potentials are able to reproduce the experimental {\it pp} and {\it pn} Wolfenstein amplitudes $a$ and $c$, which are used to compute the central and the spin-orbit part, respectively, of the $NN$ $t$ matrix; at 200 MeV the set of potentials with lower cutoffs $\Lambda$ fail to reproduce empirical data. 

The results for the observables of elastic proton scattering on $^{16}$O computed at an energy of 200 MeV are displayed in Figure~3.  The chosen energy value is rather high, in order to enlarge the differences between the different potentials, that increase with increasing energy and scattering angle, but within the limit of applicability for chiral potentials. For energies larger than $200$ MeV the agreement between the results from chiral potentials and data gets worse and it is plausible to believe that chiral perturbation theory  is no longer applicable \cite{ChOP}. 
The results of Figure~3 show that OPs obtained with the lower cutoffs (EM-450 and EGM-450) are unable to reproduce  the experimental data, while the other sets of potentials well describe the experimental cross sections and the analyzing power, that is reasonably described not only for small scattering angles but also for values larger than the minimum value up to about $45$ degrees. We note that all sets of potentials give close results for lower proton energies.  

On the basis of these results we can draw two conclusions: 
1) OPs with lower cutoffs cannot reproduce experimental data at energies close to $200$ MeV.
2) There is no appreciable difference in using $500$ or $600$ MeV as LS cutoffs, even if the EM-$600$ and EGM-$600$ potentials seem to have a slightly better agreement with empirical data, in particular looking at polarization observables.

\section*{Acknowledgments}

I thank all the people who contributed to this work, in particular Andrea Meucci and Matteo Vorabbi. This paper is dedicated to the memory of my teacher and friend Franco Capuzzi who taught me so much about the optical potential.

\end{document}